\documentclass[aps,showpacs,twocolumn,longbibliography,superscriptaddress]{revtex4-1}
\pdfoutput=1
\usepackage[utf8]{inputenc}
\usepackage[english]{babel}
\usepackage[T1]{fontenc}
\usepackage{amsmath}
\usepackage{xcolor}
\colorlet{myPurple}{blue!40!red}
\colorlet{myPurplee}{blue!10!red}
\colorlet{myCyan}{cyan!60!gray}
\colorlet{myRed}{red!66!black}
\usepackage{tikz}
\usepackage{pgfplots}
\pgfplotsset{compat=1.14}
\usepackage[colorlinks=true,citecolor=myRed,urlcolor=myRed,linkcolor=myRed]{hyperref}
\usepackage[normalem]{ulem}
\usepackage{exscale}
\usepackage{bbm}
\usepackage{graphicx}
\usepackage{amsmath}
\usepackage{latexsym}
\usepackage{amsfonts}
\usepackage{amssymb}
\usepackage{times}
\usepackage[T1]{fontenc}
\usepackage{amsthm}
\usepackage{enumerate}
\usepackage{bbold}
\usepackage{color}
\usepackage{nicefrac}%braket stuff
\newcommand{\sket}[1]{{\ensuremath{\lvert#1\rangle}}}
\newcommand{\lket}[1]{{\ensuremath{\left\lvert#1\right\rangle}}}
\newcommand{\ket}[1]{\if@display\lket{#1}\else\sket{#1}\fi}

\newcommand{\sbra}[1]{{\ensuremath{\langle#1\rvert}}}
\newcommand{\lbra}[1]{{\ensuremath{\left\langle#1\right\rvert}}}
\newcommand{\bra}[1]{\if@display\lbra{#1}\else\sbra{#1}\fi}

\newcommand{\sbraket}[2]{{\ensuremath{\langle#1\rvert#2\rangle}}}
\newcommand{\lbraket}[2]{{\ensuremath{\left\langle#1\!\left\rvert\vphantom{#1}#2\right.\!\right\rangle}}}
\newcommand{\braket}[2]{\if@display\lbraket{#1}{#2}\else\sbraket{#1}{#2}\fi}

\newcommand{\sketbra}[2]{{\ensuremath{\lvert #1\rangle\!\langle #2\rvert}}}
\newcommand{\lketbra}[2]{{\ensuremath{\left\lvert #1\right\rangle\!\!\left\langle #2\right\rvert}}}
\newcommand{\ketbra}[2]{\if@display\lketbra{#1}{#2}\else\sketbra{#1}{#2}\fi}
\usepackage{tcolorbox}
\usepackage{mathtools}

\newcommand{\proj}[1]{\ketbra{#1}{#1}}

\newcommand{\tr}{\textrm{Tr}}
\newcommand{\idd}{\mathds{1}}

\newcommand{\rA}{\text{A}}
\newcommand{\rB}{\text{B}}

\newcommand{\M}{\mathsf{M}}

\usepackage{tikz}
\usepackage{lipsum}
\theoremstyle{plain}
\usepackage{graphicx}
\usepackage{bm}
\usepackage{dsfont}
\usepackage{tikz}
\usepackage[T1]{fontenc}
\usepackage{amsthm}
\usepackage{array}
\usepackage{amssymb}
\usepackage{amsfonts}
\usepackage{cancel}
\usepackage[toc,page]{appendix}
\usepackage{multirow}
\usepackage{color}
\usepackage{calrsfs}
\usetikzlibrary{backgrounds,decorations.pathreplacing,calc}

\usepackage{tkz-euclide}
\DeclareMathAlphabet{\mathcal}{OMS}{cmsy}{m}{n}
\begin{document}

\title{Quantum nonlocality in presence of strong measurement dependence}

\author{Ivan \v{S}upi\'{c}}
\affiliation{Département de Physique Appliquée, Université de Genève, 1211 Genève, Switzerland}
\affiliation{CNRS, LIP6, Sorbonne Universit\'{e}, 4 place Jussieu, 75005 Paris, France}
\email{ivan.supic@lip6.fr}
\author{Jean-Daniel Bancal}
\affiliation{Département de Physique Appliquée, Université de Genève, 1211 Genève, Switzerland}
\affiliation{Université Paris-Saclay, CEA, CNRS, Institut de Physique Théorique, 91191, Gif-sur-Yvette, France}
\author{Nicolas Brunner}
\affiliation{Département de Physique Appliquée, Université de Genève, 1211 Genève, Switzerland}

\date{\today}

\begin{abstract}
It is well known that the effect of quantum nonlocality, as witnessed by violation of a Bell inequality, can be observed even when relaxing the assumption of measurement independence, i.e.~allowing for the source to be partially correlated with the choices of measurement settings. But what is the minimal amount of measurement independence needed for observing quantum nonlocality? Here we explore this question, and consider models with strong measurement-dependent locality, where measurement choices can be perfectly determined in almost all rounds of the Bell test. Yet, we show that quantum nonlocality can still be observed in this scenario, which we conjecture is minimal within the framework we use. We also discuss potential applications in randomness amplification. 
\end{abstract}

\maketitle

\section{Introduction}

Quantum theory allows for strong nonlocal correlations is witnessed by the violation of Bell inequalities \cite{Bell}. This phenomenon, known as quantum nonlocality, has been verified experimentally in a variety of physical platforms, see e.g.~\cite{Aspect,Tittel,Rowe,Christensen,Hensen,Giustina,Shalm}, and represent the basic resource for quantum information processing in the so-called device-independent setting. 

A Bell test typically involves a number of distant observers (say Alice and Bob) performing randomly chosen local measurements on a shared physical system. Bell inequalities are then usually derived under two assumptions (see e.g.~\cite{reviewBell}): (i) the choices of local measurements made by each party are independent from the source distributing the shared physical resource, and (ii) the measurement output of each party is determined solely by their input and a shared local (hidden) variable. Clearly, the first assumption is important, since if the source could know a priori the choice of measurement setting for each round, then all possible correlations can be reproduced by a classical model. Therefore, the observation of (quantum) nonlocality relies on the assumption that the choice of measurement settings cannot be perfectly determined at the source. In other words, there must be some level of measurement independence. But then, how much measurement independence is required for observing quantum nonlocality?

This question has attracted broad attention in recent years, and has been discussed following various approaches, see e.g.~\cite{Hall_2010,Hall_2011,Barrett_2011,Banik_2012,Koh_2012,Thinh_2013,Tan,Chaves_2015,Friedman_2019}. A notable approach is that of P\"utz et al.~\cite{Puetz_2014,putz2016measurement} who presented a general framework for addressing these questions. In particular, they derived Bell inequalities for testing local models with measurement dependent locality (MDL), i.e.~where the above assumption (i) is relaxed. Remarkably, violation of such inequalities are possible for any level of measurement dependence, demonstrating that quantum nonlocality can be observed even if measurement settings can be almost perfectly determined by the source. Hence quantum theory allows for measurement-dependent nonlocality. These ideas have been tested experimentally~\cite{Aktas_2015}, and shown to be relevant for the task of randomness amplification~\cite{Kessler} and network quantum nonlocality \cite{Supic}

A natural question at this point is whether the above works have identified the minimal requirements in terms of measurement independence, or if quantum nonlocality could in fact be demonstrated considering an even stronger form of measurement dependence. 

This is the main motivation behind the present work. Our main result is that quantum theory allows in fact for a much stronger form of measurement-dependent nonlocality. Specifically, we consider the approach of P{\"u}tz et al.~\cite{Puetz_2014,putz2016measurement}, who showed that the measurement inputs can be almost perfectly determined in each round of the Bell test. This means that in every round, the inputs feature some (possibly arbitrarily small) level of randomness with respect to the source. In contrast, we consider models where, in almost all rounds, the source can fully determine the measurement inputs. Hence we only require the presence of rounds where the inputs cannot be perfectly determined, and that these rounds have a non-vanishing probability to occur. We then show that quantum correlations are still incompatible with respect to these models. Our work thus shows that the requirement for demonstrating quantum nonlocality can still be considerably relaxed compared to previously known results. We conclude with a discussion of further questions and applications for randomness amplification.

\section{General setting}

We consider first bipartite Bell tests involving two non-communicating agents, Alice and Bob. Alice can choose between several measurements, labeled with $x \in \{0,1,\cdots,m_\rA - 1\}$. Her outcomes are labelled with $a \in \{0,1,\cdots, n_\rA -1\}$. For Bob the measurement choice label is $y \in \{0,1,\cdots, m_\rB-1\}$ and the corresponding outcome $b \in \{0,1,\cdots, n_\rB-1\}$. The experiment is characterized by the set of conditional probabilities of obtaining the pair of outcomes $a,b$ when the chosen inputs are $x$ and $y$:
\begin{equation}
   \mathcal{P} = \{p(a,b|x,y)\}_{a,b,x,y},
\end{equation}
usually called a \textit{behaviour}. Without loss of generality the correlation probabilities can be written as 
\begin{align}\label{corr}
    p(a,b|x,y) &= \int_\lambda d\lambda p(\lambda) p(x,y|\lambda)p(a,b|x,y,\lambda), 
    %&= \int_\lambda d\lambda p(\lambda|x,y)p(a|x,y,\lambda)p(b|a,x,y,\lambda),
\end{align}
where $p(\lambda)$, $p(x,y|\lambda)$  and $p(a,b|x,y,\lambda)$ are valid probability distributions. The classical variable $\lambda$ thus encodes the correlations between various measurement outputs.

Bell \cite{Bell} formalized the concept of an LHV model via the following requirements. (i) Measurement independence: the variable $\lambda$ is completely independent from the inputs $x$ and $y$, i.e. 
\begin{align} \label{MeasIndep}
    p(x,y|\lambda)=p(x,y) \quad \forall \lambda
\end{align} 
(ii) Bell locality: locally, each output is determined solely by the input and the shared variable $\lambda$, i.e. 
\begin{align}\label{Bell_local}
p(a,b|x,y,\lambda)=p(a|x,\lambda)p(b|y,\lambda) \quad \forall x,y,a,b,\lambda.
\end{align} 
A behaviour is then termed (Bell) local if it admits a decomposition of the form: 
\begin{equation}\label{local}
    p(a,b|x,y) = \int_\lambda d\lambda p(\lambda)p(a|x,\lambda)p(b|y,\lambda).
\end{equation}
We denote such a behaviour as $\bar{\mathcal{P}}_{\textrm{loc}}$ with corresponding probabilities $p_{\textrm{loc}}(a,b|x,y)$. All such behaviours satisfy Bell inequalities, which are based on linear functional of the joint probabilities, and take the form:
\begin{equation}\label{bi}
    \mathcal{I} \equiv  \sum_{a,b,x,y}\omega_{a,b,x,y}p_{\textrm{loc}}(a,b|x,y) \leq \beta_{\mathsf{loc}},
\end{equation}
where $\omega_{a,b,x,y}$ are real coefficients and $\beta_{\mathsf{loc}}$ is termed the local bound of the Bell inequality.

Violation of such a Bell inequality in presence of measurement independence implies nonlocality. This is possible in quantum theory, by performing well-chosen sets of local measurements on a shared entangled state $\varrho^{\rA\rB}$. In general, a quantum behaviour takes the form: 
\begin{equation}
    p(a,b|x,y) = \tr\left[\M_{a|x}\otimes\M_{b|y}\varrho^{\rA\rB}\right],
\end{equation}
where the sets of operators $\{\M_{a|x}\}$ and $\{\M_{b|y}\}$ represent the local measurements of Alice and Bob respectively.

Before continuing, let us introduce two additional bounds for a Bell inequality, relevant for our work. First, the maximal value of a Bell expression (the functional defined in \eqref{bi}) in quantum theory (optimized over all quantum behaviours) is called the quantum bound and denoted by $\beta_\textrm{q}$. Second, the maximal value of a Bell expression (considering any valid joint probability distribution) is called the algebraic bound and denoted by $\beta_\textrm{alg}$.

\begin{figure}
    \centering
    \includegraphics[width=1\columnwidth]{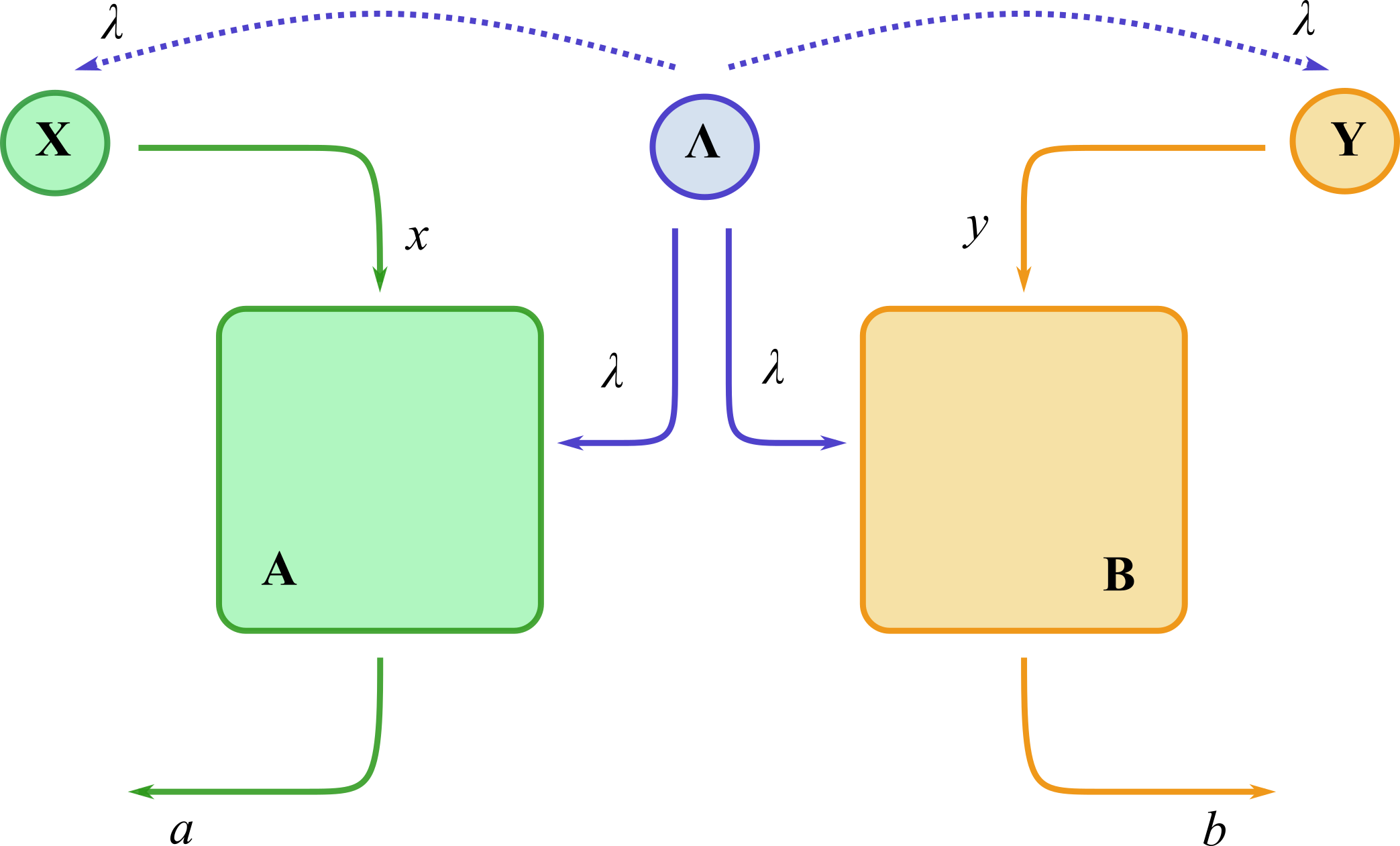}
    \caption{We consider a Bell test with relaxed measurement independence. That is, the source (i.e.~the classical variable $\Lambda$) can be correlated to the choice of measurement inputs (variables $X$ and $Y$) of both parties. We show that quantum nonlocality can exhibit strong measurement-dependent nonlocality, i.e.~the quantum predictions cannot be explained by a local model even if the source $\Lambda$ completely determines both inputs in almost all rounds of the experiment.}
    \label{fig:MDNL}
\end{figure}

Let us now come back to the assumptions behind local models. Our focus here is on the assumption of measurement independence. From a fundamental perspective, it is natural to ask whether this assumption could be partially relaxed. That is, could one still demonstrate quantum nonlocality when considering partial correlations between the source (i.e.~the shared variable $\lambda$) and the measurement inputs $x$ and $y$. Refs \cite{Thinh_2013,Puetz_2014,putz2016measurement} introduced a general framework to tackle this question. In particular, they proposed to relax the usual measurement independence condition to the following: 
\begin{equation}\label{condition}
\xi \leq  p(x,y|\lambda) \leq \eta, \qquad \forall x,y,\lambda,
\end{equation}
where $\xi > 0$ and consequently $\eta < 1$, as $\sum_{x,y}p(x,y|\lambda) = 1$. Hence the measurement inputs are no longer independent from the shared variable $\lambda$. Importantly the inputs can still not be perfectly determined from $\lambda$, as ensured by the condition $\xi > 0$. More generally, one can set given values to the parameters $\xi$ and $\eta$ which then quantify the level of measurement independence.
%Note however that the distribution of inputs can be determined by the observers, and will be chosen here to be uniform, (MOVE???) \emph{i.e.}
%
%\begin{equation}\label{eqprobinputs}
%    p(x,y) = \frac{1}{m_Am_B}, \qquad \forall x,y.
%\end{equation}
%
One can then define classes of local models with relaxed measurement independence, satisfying the condition of Bell locality (Eq. \eqref{Bell_local}) and the partial measurement independence condition in Eq. \eqref{condition} for given values of $\xi$ and $\eta$. Ref. \cite{Puetz_2014} then showed how to construct Bell inequalities for these models. A key idea consists in considering joint distributions $p(a,b,x,y)$ instead of conditional ones. Starting from a standard Bell inequality, as in \eqref{bi}, one can then construct a novel Bell inequality of the form
\begin{align}\nonumber
    \tilde{\mathcal{I}}_{\xi,\eta} \leq \xi\eta\beta_{\textrm{loc}},
\end{align}
where $\tilde{\mathcal{I}}_{\xi,\eta}$ is a linear functional of the joint probabilities $p(a,b,x,y)$ (with coefficients obtained from the original Bell functional).%, and $\tilde{\mathcal{P}}_{\textrm{loc}}$ denotes the set of local joint behaviours $\{p(a,b,x,y)\}_{a,b,x,y}$. 

Violation of such a Bell inequality under the condition Eq.~\eqref{condition} implies so-called ``measurement-dependent nonlocality''. Remarkably, P\"utz et al. \cite{Puetz_2014} showed that quantum theory allows for measurement-dependent nonlocality even for an arbitrarily small level of measurement independence, i.e.~for any $\xi>0$. Their example is connected to the quantum correlations arising from the well-known Hardy paradox \cite{Hardy}.

\section{Strong measurement-dependent nonlocality}

We are now in position to present our main result, namely that quantum correlations can in fact exhibit an even stronger form of measurement-dependent nonlocality. To be more precise, let us first notice that in the approach of P\"utz et al. \cite{Puetz_2014} reviewed above, the relaxed measurement independence condition \eqref{condition} must hold in every round of the Bell experiment. That is, in every round, the inputs cannot be perfectly determined from the shared variable $\lambda$.

Below we consider local models where this constraint is almost fully relaxed. Specifically, we consider two types of rounds. First, we have rounds where the shared variable perfectly determines the inputs (i.e.~full measurement dependence). These rounds are denoted by the set of shared variables $\Lambda''$. Second, we have rounds with relaxed measurement independence, i.e.~the condition \eqref{condition} applies for a fixed value of $\xi$ and $\eta$. This set of rounds is denoted by $\Lambda'$. Let us characterize the relative probabilities of these sets:
\begin{align}\label{weakcondition}
    \int_{\Lambda'} d\lambda p(\lambda) = q, \qquad \int_{\Lambda''} d\lambda p(\lambda) = 1-q.
\end{align}

Note that we must require that $q>0$ in order to possibly observe nonlocality. Below, we show how to construct a relevant Bell inequality, starting from any standard Bell inequality \eqref{bi}, in order to test the above model, for any values of $q>0$ and $\xi>0$. Moreover, 
we exhibit an example of such a Bell inequality, where quantum correlations lead to a Bell violation for any values of $q>0$ and $\xi>0$. We say that such correlations feature strong measurement-dependent nonlocality.

%Consider now the Bell inequality whose maximal quantum violation $\beta_{\textrm{q}}$ is equal to its algebraic bound $\beta_\textrm{alg}$. In the bipartite scenario such is the inequality corresponding to the Mermin-Peres magic square game, and in the tripartite scenario the Mermin inequality. 

Consider now an arbitrary Bell inequality~\eqref{bi}. As in~\cite{Puetz_2014} we denote $\omega_{a,b,x,y}^+ = \omega_{a,b,x,y}$ if $\omega_{a,b,x,y} > 0$ and $\omega_{a,b,x,y}^- = \omega_{a,b,x,y}$ if $\omega_{a,b,x,y} > 0$. Then the Bell inequality has the form 
\begin{align}\label{bialt}
    \mathcal{I} &\equiv  \sum_{a,b,x,y}\left[\omega_{a,b,x,y}^+p_{\textrm{loc}}(a,b|x,y) - \omega_{a,b,x,y}^-p_{\textrm{loc}}(a,b|x,y)\right]\\ &\leq \beta_{\mathsf{loc}}\nonumber.
\end{align}
The quantum bound of the inequality, equal to $\beta_\mathrm{q}$ can be decomposed as $\beta_\mathrm{q} = \beta_\mathrm{q}^+ - \beta_\mathrm{q}^-$, where we introduced notation
\begin{align}\label{betaq+}
    \beta_\mathrm{q}^+ &= \sum_{a,b,x,y}\omega_{a,b,x,y}^+p_\mathrm{q}(a,b|x,y)\\ \label{betaq-}
    \beta_\mathrm{q}^- &= \sum_{a,b,x,y}\omega_{a,b,x,y}^-p_\mathrm{q}(a,b|x,y),
\end{align}
and $\{p_q(a,b|x,y)\}$ is a behaviour reaching the quantum bound. The measurement-dependent Bell inequality from~\cite{Puetz_2014} reads
\begin{align}\label{mdBi}
    &\tilde{\mathcal{I}}_{\xi,\eta} = \sum_{a,b,x,y} \bigg(\xi\omega_{a,b,x,y}^+p(a,b,x,y) - \eta\omega_{a,b,x,y}^-p(a,b,x,y)\bigg).
\end{align}    
By separating the sets $\Lambda'$ and $\Lambda''$ and taking into account the decomposition of local joint probabilities we get    
\begin{widetext}
\begin{align*}
\tilde{\mathcal{I}}_{\xi,\eta}    &= \sum_{a,b,x,y} \bigg(\xi\omega_{a,b,x,y}^+(1-q)p(a,b,x,y) - \eta\omega_{a,b,x,y}^-(1-q)p(a,b,x,y)\bigg) + \\ 
    &\qquad + \sum_{a,b,x,y}\left(\xi\omega_{a,b,x,y}^+q\int_{\Lambda'} d\lambda \frac{p(\lambda)}{q}p(x,y|\lambda)p(a|x,\lambda)p(b|y,\lambda)-\eta\omega_{a,b,x,y}^-q\int_{\Lambda'} d\lambda \frac{p(\lambda)}{q}p(x,y|\lambda)p(a|x,\lambda)p(b|y,\lambda)\right).
\end{align*}
\end{widetext}
 The second line is simply the renormalized measurement-dependent inequality for the local variables satisfying the condition \eqref{condition}. Hence its local bound is $q\xi\eta\beta_\mathrm{loc}$, where $q$ comes from the normalization of the hidden variable probability distribution. In the first line, as the condition \eqref{condition} is not satisfied, the second term can be equal to $0$. Observe now that the algebraic bound is reached if for all $a,b,x,y$ such that $\omega_{a,b,x,y} < 0$ it holds $p(a,b,x,y) = 0$. As the condition~\eqref{condition} is not satisfied, the optimal value of the first term in the equation is achieved when $p(x,y|\lambda) = 1$ for all $x,y$ and $\lambda \in \Lambda''$. Hence the maximal value of the first line is 
\begin{equation*}
    \sum_{a,b,x,y} \xi\omega_{a,b,x,y}^+(1-q)p(a,b,x,y) = \xi(1-q)\tilde{\beta}_{\mathrm{alg}},
\end{equation*}
where $\tilde{\beta}_{\mathrm{alg}}$ is the algebraic bound of the inequality $\sum_{a,b,x,y}\left[\omega_{a,b,x,y}^+p_{\textrm{loc}}(a,b,x,y) - \omega_{a,b,x,y}^-p_{\textrm{loc}}(a,b,x,y)\right]$.
This implies that the local bound for the strong measurement-dependent LHV models is
\begin{equation}\label{smdloc}
    \beta_\mathrm{loc}^{smd} \equiv (1-q)\xi\tilde{\beta}_\mathrm{alg} + q\xi\eta{\beta}_\mathrm{loc}.
\end{equation}
Thus, for hidden variables satisfying condition \eqref{condition} when $\lambda \in \Lambda'$ and $\Lambda'$ satisfying condition \eqref{weakcondition} the bound of the measurement-dependent Bell inequality becomes:
\begin{equation}
    \mathcal{I}^{md} \leq \beta_\mathrm{loc}^{smd}.
\end{equation}

The next part consists in evaluating the quantum bound of the expression $\tilde{\mathcal{I}}_{\xi,\eta} $. For this, it is useful to know the distribution of measurement inputs, which can be observed in the Bell test. From now on, let us assume that all possible input pairs appear with equal probabilities, \emph{i.e.} $p(x,y) = 1/(m_Am_B)$. From the definitions~\eqref{betaq+} and~\eqref{betaq-}, we get the quantum bound
\begin{equation}\label{smdq}
    \beta_\mathrm{q}^{smd} = \frac{1}{m_Am_B}\left(\xi\beta_\mathrm{q}^+ - \eta\beta_\mathrm{q}^-\right) \,.
\end{equation}
Strong measurement-dependent nonlocality is then observed when $\beta_\mathrm{q}^{smd}>\beta_\mathrm{loc}^{smd}$. Below we discuss when such a Bell violation is possible.

For most Bell inequalities, we have that $\beta_\mathrm{q} < \beta_\mathrm{alg}$. It follows that one can detect strong measurement-dependent nonlocality for some limited range of parameters of the model. Notably, one will require that $q>q_0$, where $q_0$ is a strictly positive constant. This means that the rounds where the local variable cannot perfectly determine the inputs have a certain minimal probability to occur. This is for example the case for the quantum correlations presented in Ref. \cite{Puetz_2014}.

%There exist however Bell inequalities for which $\beta_\mathrm{q} = \beta_\mathrm{alg}$; a feature also referred to as quantum ``pseudo-telepathy''.

It is however possible to construct quantum correlations that are incompatible with any local model with strong MDL, i.e.~considering any parameter values of $q > 0$ and $\xi > 0$. We now present an example of such quantum correlations. Our starting point are quantum nonlocal correlation arising from the Peres-Mermin (PM) magic-square game~\cite{Peres_1991,Mermin}, which have been proposed in Ref. \cite{Aolita2012}. This work also constructs a Bell inequality to detect these nonlocal correlations. The key property of this Bell inequality is that the quantum bound coincides with the algebraic one; a feature sometimes referred to as quantum ``pseudo-telepathy''.

Following Ref. \cite{Aolita2012} we consider a bipartite Bell test with three inputs per party. For each input, each party outputs two bits, denoted $a = (a_1,a_2)$ and for Bob with $b = (b_1,b_2)$, with $a_i,b_j \in\{-1,1\}$. For our purpose, we need to construct a variant of the Bell inequality of Ref. \cite{Aolita2012}. Let us first introduce two types of correlators:
\begin{align}
    \tilde{C}_{x,y}^{f(a,b)} &= \sum_{a,b}2^{\frac{f(a,b)+1}{2}}(-16)^{-\frac{f(a,b)-1}{2}}p(a,b|x,y),\\
    \bar{C}_{x,y}^{f(a,b)} &= \sum_{a,b}16^{\frac{f(a,b)+1}{2}}(-2)^{-\frac{f(a,b)-1}{2}}p(a,b|x,y).
\end{align}
While it is common to use correlators taking values in the interval $ [-1,1]$ for Bell inequalities, note that the above modified correlators take values in the following intervals: $-16 \leq \tilde{C}_{x,y}^{f(a,b)}\leq 2$ and $-2 \leq \bar{C}_{x,y}^{f(a,b)}\leq 16$ (it will become clear below why we need this). 
%For a given pair of inputs, the correlator $\tilde{C}_{x,y}^{f(a,b)}$ takes the double sum of probabilities to obtain outputs $a,b$ such that $f(a,b) = 1$ and subtracts the sum of probabilities to obtain all other outputs multiplied by $16$, hence  . Correlator $\bar{C}_{x,y}^{f(a,b)}$ does the opposite: sum of probabilities to obtain outputs $a,b$ such that $f(a,b)=1$ is multiplied by $16$, and from it is subtracted the sum of probabilities for obtaining all other outputs, multiplied by $2$, hence $-2 \leq \bar{C}_{x,y}^{f(a,b)}\leq 16$. The standard Bell inequality we introduce here is a modified version of the Peres-Mermin magic-square game~\cite{Peres_1991,Mermin}:
%
We now construct the following Bell expression
\begin{align} \nonumber
    I &= \tilde{C}_{1,1}^{a_1b_1} + \tilde{C}_{1,2}^{a_2b_1} + \tilde{C}_{2,1}^{a_1b_2} 
    + \tilde{C}_{2,2}^{a_2b_2} +  \tilde{C}_{1,3}^{a_1a_2b_1} + \\ \label{modifPM} &+ \tilde{C}_{2,3}^{a_1a_2b_2} 
    + \tilde{C}_{3,3}^{a_1a_2b_1b_2} + \tilde{C}_{3,1}^{a_1b_1b_2} - \bar{C}_{3,2}^{a_2b_1b_2} \,,
\end{align}
with algebraic bound $\beta_{alg}=18$.  Using the quantum strategy based on the PM magic-square game as in \cite{Aolita2012}, we get a nonlocal distribution that reaches $I=18$, hence we have that $\beta_{q}=\beta_{alg}$.

Now the important property for our purpose is that the local bound of $I$ is $\beta_{loc}=0$ (hence the use of modified correlators), which simply follows from inspection over all deterministic strategies. 

%as each of first eight terms can achieve maximal value $2$, while for the last term the minimal value is $-2$. The quantum bound for this inequality coincides with the algebraic, and it can be attained by using exactly the same strategy used for winning the well-known Peres-Mermin magic square game with the highest probability.  The notable difference, which is convenient for our purpose, is that the local bound is now $\beta_{loc}=0$. 

\begin{figure}
    \centering
    \includegraphics[width=1\columnwidth]{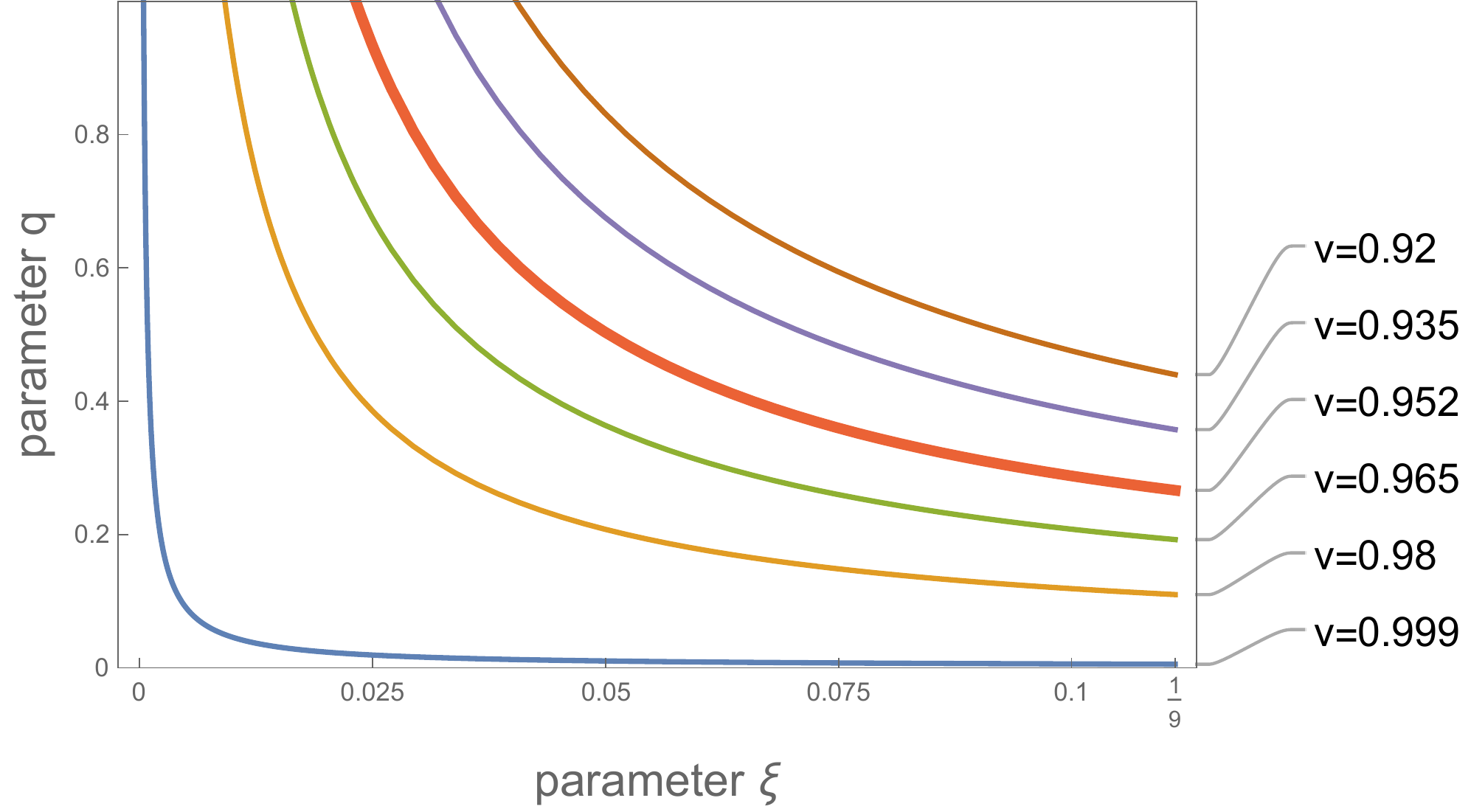}
    \caption{The graph depicts nonlocality in the presence of strong measurement-dependence attained by the states $\proj{\phi_v}= v\proj{\phi_+} + (1-v)\idd/16$, where $\ket{\phi_+} = \sum_{j=0}^3\ket{jj}/2$. The measurements correspond to those used to obtain the maximal violation of the Mermin-Peres magic square game. For given visibility $v$ nonlocality is observed if the values of parameters $\xi$ and $q$ belong to the region above the corresponding curve. The parameter $\eta$ is taken to be equal to $1/9 + \xi$. The red bold curve corresponds to a visibility comparable to the experiment of Ref.~\cite{Aolita2012}, based on a hyper-entangled state..}
    \label{fig:Visibility}
\end{figure}

We can now apply the method outlined above to adapt this Bell inequality to the measurement-dependent scenario, namely 
\begin{align} \nonumber
    \tilde{\mathcal{I}}_{\xi,\eta} &= \tilde{C}_{1,1,\xi,\eta}^{a_1b_1} + \tilde{C}_{1,2,\xi,\eta}^{a_2b_1} + \tilde{C}_{2,1,\xi,\eta}^{a_1b_2} 
    + \tilde{C}_{2,2,\xi,\eta}^{a_2b_2} +  \tilde{C}_{1,3,\xi,\eta}^{a_1a_2b_1} + \\ \label{MDmodifPM} &+ \tilde{C}_{2,3,\xi,\eta}^{a_1a_2b_2}
    + \tilde{C}_{3,3,\xi,\eta}^{a_1a_2b_1b_2} + \tilde{C}_{3,1,\xi,\eta}^{a_1b_1b_2} - \bar{C}_{3,2,\xi,\eta}^{a_2b_1b_2},
\end{align}
where $\tilde{C}_{x,y,\xi,\eta}^{f(a,b)}$ and $\bar{C}_{x,y,\xi,\eta}^{f(a,b)}$ are now  weighted correlators of joint distributions, defined as
\begin{align}
    \label{WCMD}
    \tilde{C}_{x,y,\xi,\eta}^{f(a,b)} &= \sum_{a,b}\xi 2^{\frac{f(a,b)+1}{2}}(-16\eta)^{-\frac{f(a,b)-1}{2}}p(a,b,x,y),\\
    \bar{C}_{x,y,\xi,\eta}^{f(a,b)} &= \sum_{a,b}\eta 16^{\frac{f(a,b)+1}{2}}(-2\xi)^{-\frac{f(a,b)-1}{2}}p(a,b,x,y).
\end{align}
Notice that for $ \tilde{\mathcal{I}}_{\xi,\eta}$ we have $\tilde{\beta}_{\mathrm{alg}}=2$, hence from eq.~\eqref{smdloc} we obtain the local bound $\beta_\mathrm{loc}^{smd} = 2(1-q)\xi$, while the quantum bound is $\beta_\mathrm{q}^{smd}=2\xi$. Hence we obtain a quantum violation for any parameter values $q>0$ and $\xi>0$. On Fig.~\ref{fig:Visibility} we show for which values of parameters $\xi$ and $q$ nonlocality with strong measurement dependence can be observed if noisy states are used to violate the Mermin-Peres magic square game. 

In the Appendix, we also present a similar construction based on the Mermin inequality, considering a tripartite Bell test.

\section{Connection to randomness amplification}

While the question of observing quantum nonlocality under conditions of relaxed measurement independence is of rather fundamental nature, these ideas also have implications from a more applied perspective. Indeed, one prominent application of quantum nonlocality is the task of generating randomness. This process is known as ``device-independent'' quantum random number generation (DI-QRNG), which explores the possibility to generate certified random numbers under minimal assumptions on the devices; see e.g.~\cite{Acin_2016} for a recent review. 

The most relevant aspect of DI-QRNG here is the task of DI randomness amplification~\cite{Colbeck_2012}. The idea is to perform a Bell test using imperfect sources of randomness to generate the inputs, and being able to certify more randomness in the outputs, hence amplifying randomness via nonlocal correlations. Previous works in this line have demonstrated randomness amplification considering so-called Santha-Vazirani sources for inputs~\cite{Colbeck_2012,Gallego_2013,miller2016robust,Brand_o_2016,Ramanathan_2016}. Interestingly, randomness amplification can be directly connected to the measurement-dependent nonlocality~\cite{Kessler}, specifically to the approach of Ref.~\cite{Puetz_2014}, which leads to protocols with improved performance. 

Given this strong connection, we believe that our approach can have direct consequences in the context of randomness amplification. In particular, our results should translate into stronger forms of randomness amplifications, going beyond Santha-Vazirani sources. Another interesting question is to understand how our approach connects to the work of Ref.~\cite{Thinh_2013}, where min-entropy sources are considered.

\section{Discussion}

We explored the question of observing quantum nonlocality considering local models with relaxed measurement independence. Using the approach developed by P\"utz et al. \cite{Puetz_2014}, we showed that the requirements in terms of measurement independence can still be considerably relaxed, and presented an example of quantum correlations featuring nonlocality in presence of strong measurement dependence. This suggests that nonlocality could be demonstrated with strings of measurement settings having only a very small amount of global min entropy.

The main open question is arguably whether our work has now reached the absolute minimal requirements in terms of measurement independence allowing for quantum nonlocality. Of course, this question is only meaningful within a certain framework. For the approach developed in P\"utz et al. \cite{Puetz_2014}, we believe that our work has identified the minimal requirements. An interesting question is now to see how this result may impact applications in randomness amplification. 

Finally, our work has also interesting consequences from the point of view of network nonlocality. Following the construction of Ref. \cite{Supic}, one can embed our example of strong measurement-dependent quantum nonlocality in a triangle network without inputs. In this way, one obtains a quantum distribution that cannot be reproduced classically even when the three sources of the triangle network are perfectly correlated in almost all rounds of the Bell test.

\emph{Acknowledgements.---} We acknowledge financial support from the Swiss National Science Foundation (project 2000021\_192244/1 and NCCR SwissMAP) and ERC Starting grant QUSCO.

\bibliographystyle{unsrturl}
\bibliography{main.bbl}

\section{Appendix}

In this Appendix, we show another example of quantum correlations featuring strong MDL for any parameter values $q>0$ and $\xi>0$. This is based on a tripartite Bell test, involving a third party, Charlie, whose input is denoted $z$ and output $c$. The main idea is to start from the Mermin Bell inequality \cite{Mermin} which also has the property that the quantum bound is equal to the algebraic one. The optimal quantum strategy is that of the famous GHZ paradox.

As our construction in the main text, we must however first modify the Mermin inequality in a form that is suitable for our purpose. This implies using suitable definitions for correlators. For a given set of inputs $x,y,z$ we define the weighted correlator:
\begin{equation}
    \label{WC}
    \tilde{C}_{x,y,z} = \sum_{a,b,c}3^{a\oplus b \oplus c\oplus 1}(-1)^{a\oplus b \oplus c}p(a,b,c|x,y,z),
\end{equation}
with $a,b,c \in \{0,1\} $. We observe that $-1\leq \tilde{C}_{x,y,z} \leq 3$. From this, we now construct an alternative version of the Mermin Bell inequality, namely
\begin{equation}
    \label{WCineq}
    \tilde{I} = \tilde{C}_{0,0,0} - \tilde{C}_{1,1,0} - \tilde{C}_{1,0,1} - \tilde{C}_{0,1,1}. 
\end{equation}
The quantum bound for this inequality is $\tilde{I}=4$, which can be attained by considering a three-qubit GHZ state and each party performing local measurements in the Pauli $X$ and $Y$ basis (similarly as in the well-known GHZ paradox, giving maximal violation of the Mermin Bell inequality). It turns out that this quantum bound coincides with the algebraic bound of the inequality, i.e.~$\beta_{q}=\beta_{alg}$, similarly to the standard Mermin inequality. The notable difference, which is convenient for our purpose, is that the local bound is now $\beta_{loc}=0$. 

We can now apply the method outlined above to adapt this inequality to the measurement-dependent scenario, namely
\begin{equation}
    \label{WCMDineq}
    \tilde{I}^{md} = \tilde{C}^{\eta,\xi}_{0,0,0} - \tilde{C}^{\eta,\xi}_{1,1,0} - \tilde{C}^{\eta,\xi}_{1,0,1} - \tilde{C}^{\eta,\xi}_{0,1,1}, 
\end{equation}
where $\tilde{C}^{\eta,\xi}_{x,y,z}$ is now  a weighted correlator of joint distributions, defined as
\begin{equation}
    \label{WCMD2}
    \tilde{C}^{\eta,\xi}_{x,y,z} = \sum_{a,b,c}(3
    \eta)^{a\oplus b \oplus c\oplus 1}(-\xi)^{a\oplus b \oplus c}p(a,b,c,x,y,z).
\end{equation}
Finally, we find the local and quantum bounds for this new inequality: $\beta_\mathrm{loc}^{smd} = (1-q)\xi$ and $\beta_\mathrm{q}^{smd}=\xi$. Hence we obtain a quantum violation for any parameter values $q>0$ and $\xi>0$.

\end{document}